\def\year{2022}\relax
\documentclass[letterpaper]{article} 
\usepackage{aaai22}  
\usepackage{times}  
\usepackage{helvet}  
\usepackage{courier}  
\usepackage[hyphens]{url}  
\usepackage{graphicx} 
\usepackage{natbib}  
\usepackage{caption} 
\DeclareCaptionStyle{ruled}{labelfont=normalfont,labelsep=colon,strut=off} 
\usepackage{algorithm}
\usepackage{algorithmic}
\usepackage{tabularx}
\usepackage{subfigure}
\usepackage{multirow}
\usepackage{mathtools}
\usepackage{enumitem}
\usepackage{booktabs}
\usepackage{amsmath}
\usepackage{color}
\usepackage{pifont}
\newcommand{\heart}{\ding{170}}
\newcommand{\spade}{\ding{171}}
\newcommand{\club}{\ding{168}}

\usepackage{newfloat}
\usepackage{listings}
\floatstyle{ruled}
\newfloat{listing}{tb}{lst}{}
\floatname{listing}{Listing}

\usepackage{color}

\title{\#RoeOverturned: Twitter Dataset on the Abortion Rights Controversy}

\author {
    Rong-Ching Chang$^*$ \textsuperscript{\rm \spade},
    Ashwin Rao$^*$ \textsuperscript{\rm \heart,\club},
    Qiankun Zhong$^*$ \textsuperscript{\rm \spade},
    Magdalena Wojcieszak\textsuperscript{\rm \spade},
    Kristina Lerman \textsuperscript{\rm \heart,\club}
}
\affiliations {
    \textsuperscript{\rm \spade} University of California - Davis\\
    \textsuperscript{\rm \heart} University of Southern California\\
    \textsuperscript{\rm \club} Information Sciences Institute\\
    rocchang@ucdavis.edu, mohanrao@usc.edu, qkzhong@ucdavis.edu, mwojcieszak@ucdavis.edu, lerman@isi.edu
}

\begin{document}

\maketitle
\def\thefootnote{$*$}\footnotetext{These authors contributed equally to this work.}
\def\thefootnote{\arabic{footnote}}

\begin{abstract}

On June 24, 2022, the United States Supreme Court overturned landmark rulings made in its 1973 verdict in Roe v. Wade. The justices by way of a majority vote in Dobbs v. Jackson Women's Health Organization, decided that abortion wasn't a constitutional right and returned the issue of abortion to the elected representatives. This decision triggered multiple protests and debates across the US, especially in the context of the midterm elections in November 2022. Given that many citizens use social media platforms to express their views and mobilize for collective action, and given that online debate provides tangible effects on public opinion, political participation, news media coverage, and the political decision-making, it is crucial to understand online discussions surrounding this topic. Toward this end, we present the first large-scale Twitter dataset collected on the abortion rights debate in the United States. We present a set of $74M$ tweets systematically collected over the course of one year from January 1, 2022 to January 6, 2023. 

\end{abstract}

\section{Introduction}

The US Supreme Court's 1973 decision in the Roe v. Wade case gave women the right to terminate their pregnancy. Despite federal protection, abortion has remained a highly polarized and politically charged issue in American society. 
Studies published by Gallup \cite{gallup2010abortion,gallup2011abortion} noted that views of Democrats and Republicans on abortion rights had grown increasingly polarized since 1975 and Americans continued to be divided along pro-choice and pro-life lines.  A follow up study in 2021 \cite{gallup2021abortion} showed that Americans continue to differ on the morality of abortion with 47\% of survey takers finding abortion to be morally acceptable and 46\% disagreeing. These findings suggest that abortion has been one of the moralized political issues at the core of the so-called culture wars in the United States. Previous studies largely seem to agree on the highly polarizing nature of the abortion rights debate. ~\cite{dimaggio1996have,mouw2001culture,evans2003have,fiorina2008political,abramowitz2008polarization}. Although some scholars (e.g. \cite{fiorina2008political}) suggest that the general public's policy preferences are mostly centrist toward wedge-issues including abortion, or argue that the extent of polarization in the US society is exaggerated \cite{mouw2001culture}, other evidence suggests different patterns. For instance, analyzing over 20 years of data from the National Election Studies (NES) and General Social Survey (GSS) \cite{dimaggio1996have} found that polarization on the issue of abortion had increased in the US.  Similarly, leveraging NES and GSS data from 1970-2003 \cite{evans2003have} found unambiguous evidence of polarization with regards to abortion rights 
Public opinion data aside, the National Abortion Federation found that \cite{naf2022abortion} violent crimes and assaults directed towards abortion clinics and patients rose by a staggering $128\%$ from 54 cases in 2020 to 123 cases in 2021. 

The June 24, 2022 United States Supreme Court ruling in Dobbs v. Jackson Women's Health Organization struck down decisions made in Roe v. Wade (1973) and Casey v. Planned Parenthood (1992), removing federal protections for abortion and moving power to regulate it to the states. 

In delivering the opinion of the court, Justice Samuel Alito noted that abortion presents a profound moral issue on which Americans hold sharply contrasting views \cite{dobbs2022roe}. Given that states had enacted trigger laws \footnote{A trigger law is a law that is unenforceable but may achieve enforceability if a key change in circumstances occurs. \url{https://en.wikipedia.org/wiki/Trigger_law}} to go into effect once the Supreme Court overturned Roe v. Wade, abortion at the time of writing this article is banned in 13 states with or without exceptions for rape and incest. Other states have either completely banned abortion or provide limited access with gestational limits. Only 16 states in the US have legislation that protects access to abortion. Contrasting laws in different states of the country indicate the extent of the disagreement with respect to abortion rights. 


As the ruling leaked, protesters both in favor (pro-choice) and opposition (pro-life) of the Roe v. Wade ruling of 1973 gathered on the streets and mobilized on social media.  Slogans like ``Bans Off Our Bodies'' (Pro-Choice), ``Abortion is a Right'' (Pro-Choice), ``Equal rights for the unborn'' (Pro-life) and ``Time to Reverse Roe'' (Pro-life), echoed on the streets and online, reflecting a deeply polarized populace.  In fact, many of these slogans, expressions, and discussions took place on social media, especially on Twitter, as we detail below.  The easy availability of those data, furthermore, offers unprecedented insights to studying public opinion expression surrounding contentious political issues. 

The rise of social media platforms since the early 2000s has provided researchers with a novel avenue to study public opinion, individual expressions, and behavioral indicators. The advantage of social media datasets over survey-based measurements used in public opinion polls is the ability to feasibly assess issue positions of the larger public without biases inherent in survey self-reports, especially of contentious political issues (e.g., social desirability bias). Large-scale social media datasets have also made possible, the analysis of protest mobilization \cite{breuer2015social,steinert2017spontaneous,munn2021more}, proliferation of misinformation \cite{nikolov2020right,rao2021political,chen2021covid}, moral and emotional attitudes \cite{guo2022emotion, priniski2021mapping}, echo chambers and ideological biases on platforms \cite{barbera2015birds,wojcieszak2022most}, among other democratically relevant phenomena. Previous studies \cite{yardi2010dynamic,garimella2018political,cinelli2021echo} have also analyzed polarization with respect to wedge issues, including abortion rights in the US. The collection of large-scale social media data at pivotal points like the COVID-19 pandemic \cite{huang_xiaolei_2020_3735015, chen2020tracking}, 2020 US Presidential elections \cite{chen2022election2020,abilov2021voterfraud2020} and the Russian invasion of Ukraine \cite{chen2022tweets} helps researchers capture the pulse of an increasingly volatile world. 

The bulk of this work focuses on Twitter not only due to the relative ease of obtaining online behavioral data from the platform (relative to other platforms, such as Facebook) but also because Twitter is uniquely suited to analyzing discourse about political issues, given the platform’s important role in American politics. Twitter is a key outlet for individuals to express their political opinions and engage in political activities \cite{bestvater2022politics}. These tweets and engagement metrics are often treated as proxies of public opinion by journalists and campaign strategists \cite{mcgregor2019social}, set political agendas that politicians follow \cite{barbera2019leads}, and influence what journalists and media report \cite{nelson2019doing}. In addition, Twitter is an important channel that citizens use to get political information: 71\% of adult Twitter users in the U.S. report getting news on the site \cite{stocking2018sources}, and is also a key platform for elite expressions: almost all US politicians, pundits, and news media have a Twitter account and politicians are more active and have more followers on Twitter than on other platforms \cite{wojcieszak2022most}. 

Furthermore, Twitter users are more politically interested than the general population. In fact, Twitter stands out as one of the social media platforms with the most politically engaged users: 42\% of U.S. adults on Twitter say they use the site to discuss politics at least some of the time, and this percentage is higher among the most frequent tweeters \cite{wojcik2019sizing}. Furthermore, this politically active group of Twitter users is highly influential. They produce the majority of engagements with elite tweets \cite{wojcieszak2021echo} and political content on Twitter \cite{bestvater2022politics}, set political agendas \cite{barbera2019leads}, and are more likely to engage in political activities online and offline than the general public \cite{bestvater2022politics}. In short, Twitter is well suited to examine the prevalence and fluctuations of the discourse surrounding the overturning of Roe v. Wade. As Twitter users are not representative of the general population \cite{wojcik2019sizing}, the tweets we collect do not necessarily represent public opinion on abortion. They however, reflect the discourse of the politically inclined and societally influential citizens.

Our dataset makes possible the exploration of a variety of research directions. Some interesting avenues include but are not limited to - modeling opinion dynamics and polarization, protest mobilization, assessing emotions and moral attitudes and the influence of social bots. When abortion was first outlawed in the mid to late 1800s, abortion was stigmatized for close to a century. Given that hate speech and toxic content thrive on social media platforms \cite{founta2018large}, an analysis of such discourse becomes imperative to prevent abortion from being stigmatized again.  


In this study, we collect a large-scale Twitter data set consisting of discourse surrounding abortion rights in the United States and the recent Supreme Court verdict that overruled its 1973 Roe v. Wade ruling. 
Data collection is ongoing at the time of writing this article. In the following sections, we will describe data collection methods, descriptive statistics, and information on how to access and use the data.

\section*{Methods}
\subsection*{Overview}

Prior to collecting the tweets, we identified a set of keywords that are meant to represent the breadth and the different sides of the discourse surrounding abortion in the United States. 
We followed a three-pronged data collection strategy, discussed below. 
We began collecting tweets on June 25, 2022. 
In order to collect data prior to June 25, 2022, we leverage Twitter Academic API's Full-Archive Search \footnote{\url{https://developer.twitter.com/en/docs/twitter-api/tweets/search/api-reference/get-tweets-search-all}}. 
This gives us data from January 1, 2022 to June 24, 2022. 
We gather tweets in real-time from June 25, 2022 to January 6, 2023. 
To ensure our dataset coverage, we use the count end point from the Twitter API to compare the number of tweets in our collection versus the number of tweets identified on Twitter. We use the Full-Archive Search endpoint to recollect the date if the total number of tweets per day in our dataset is less than the number of tweets identified on Twitter. 
Our data coverage is from January 1, 2022 to January 6, 2023. Data collection is ongoing at the time of writing this article. We restrict our collection to English tweets. In all, we collected over $74M$ tweets shared by roughly $10M$ users. 

\begin{table}[h]
\centering
\begin{tabularx}{\columnwidth}{|c|X|}
\toprule
\textbf{Category}  & \textbf{Keywords}               \\ \hline
Neutral   & \#Abortion, \#roevswade, \#Roeverturned, \#roevwade, abortion, roe vs wade, roe v wade, roe overturned \\\hline
Pro-choice & \#roevwadeprotest, roe v wade protest, pro choice, pro-choice, \#prochoice, \#forcedbirth, forced birth, \#AbortionRightsAreHumanRights,  abortion rights Are Human Rights, \#MyBodyMyChoice, My Body My Choice, \#AbortionisHealthcare, abortion is healthcare, AbortionIsAHumanRight, abortion is a human right, ReproductiveHealth, Reproductive Health, AbortionRights, abortion rights,   \\\hline
Pro-life   & \#prolife, pro life, pro-life, \#EndAbortion, end abortion, \#AbortionIsMurder, Abortion Is Murder, \#LifeIsAHumanRight, Life Is A Human Right, \#ChooseLife, Choose Life, \#SaveTheBabyHumans, Save The Baby Humans, \#ValueLife, Value Life, \#RescueThePreborn, Rescue The Preborn, \#EndRoeVWade, End Roe V Wade, \#LifeMatters, Life Matters, \#MakeAbortionUnthinkable, make abortion unthinkable, \#LiveActionAmbassador, Live Action Ambassador, Abortion Is Not A Right, \#AbortionIsNotARight, \#AbortionIsRacist, Abortion is racist, anti life, \#antilife \\ \hline
\end{tabularx}
\caption{List of keywords}
\label{tab:hash}
\end{table}

\subsection*{Keyword collection strategy}
The Full-Archive Search endpoint of Twitter Academic API requires us to specify a set of keywords, on the basis of which the API returns relevant tweets. To this end, we conducted an exploratory collection to compile an exhaustive list of relevant phrases and hashtags to create a keyword set that covers abortion-focused discussions from different political stances. Henceforth, we will use the term \textit{keywords} to refer to both phrases and hashtags. We then build a systematic data collection pipeline, which we will discuss in the following subsections.  

\subsubsection{Hashtag sampling} 
We first identified a set of seed hashtags - \#roevwade, \#prolife, \#prochoice. The \#prolife hashtag is used more commonly by individuals and activists who believe that human life begins at conception and abortion ends the life of an innocent being. The \#prochoice hashtag on the other hand, is used more frequently by individuals who believe that it is a woman's choice whether or not to get an abortion and any regulation of abortion infringes their freedom. The \#roevwade is one hashtag that is used by individuals on both sides and is neutral per se. 

Leveraging these seed hashtags, we conducted an exploratory data collection using the Twitter API's recent tweet lookup endpoint \footnote{\url{https://developer.twitter.com/en/docs/twitter-api/tweets/search/api-reference/get-tweets-search-recent}}. The purpose of the exploratory data collection was to identify other hashtags that frequently co-occur with the seed hashtags. We used snowball sampling to identify hashtags that frequently co-occurred with \#prochoice, \#prolife and \#roevswade. 
 
From tweets that contain \#prochoice we identify the hashtags - \#roevwadeprotest, \#prochoice, \#forcedbirth, \#AbortionRightsAreHumanRights, \#MyBodyMyChoice, \#abortionishealthcare, \#abortionisahumanright, \#ReproductiveHealth, and \#abortionrights to frequently co-occur. These form the set of pro-choice hashtags.   From tweets that contained \#prolife, we find \#EndAbortion, \#AbortionIsMurder, \#LifeIsAHumanRight, \#ChooseLife. \#valueLife, \#rescuethepreborn, \#endroevwade, \#lifematters, \#antilife, \#makeabortionunthinkable, \#liveActionAmbassador,\#catholic, \#AbortionIsNotARight, and \#Abortionisracist to co-occur frequently. These hashtags form the set of pro-life hashtags. As \#Abortion, \#Roeverturned and \#roewade frequently co-occured with \#roevswade they form the set of neutral hashtags. 


\subsubsection{Creating keywords} 
Given that not all users on Twitter use hashtags in their tweets, we create a list of keywords based on the aforementioned hashtags by simply removing \# from the hashtags. For hashtags that comprised multiple words, we break them down to reconstruct the phrases (e.g., \#makeabortionunthinable into "make abortion unthinkable"). The final set of hashtags and phrases is shown in Table \ref{tab:hash}. In all we have 58 keywords comprised of 31 hashtags and 30 phrases. The keywords shown in Table \ref{tab:hash} form the search terms in our query. More specifically, our search query is comprised of these keywords connected together by the 'OR' operator, which tells the search endpoint to return tweets that have at least one keyword in it. The search is not case-sensitive. 



\subsubsection{Validation and removal} 
Given that the Roe v. Wade overturns was US-specific (although it was covered internationally and generated global debates) and the most frequently used keywords are all in English, we do not collect non-English tweets \footnote{According to our collection of all language from June to September 2022, over 90\% of the tweets on this topic is English. We thus removed the non-English tweets and collect only English tweets in this dataset.}.  

\begin{figure}
    {\includegraphics[width=\linewidth]{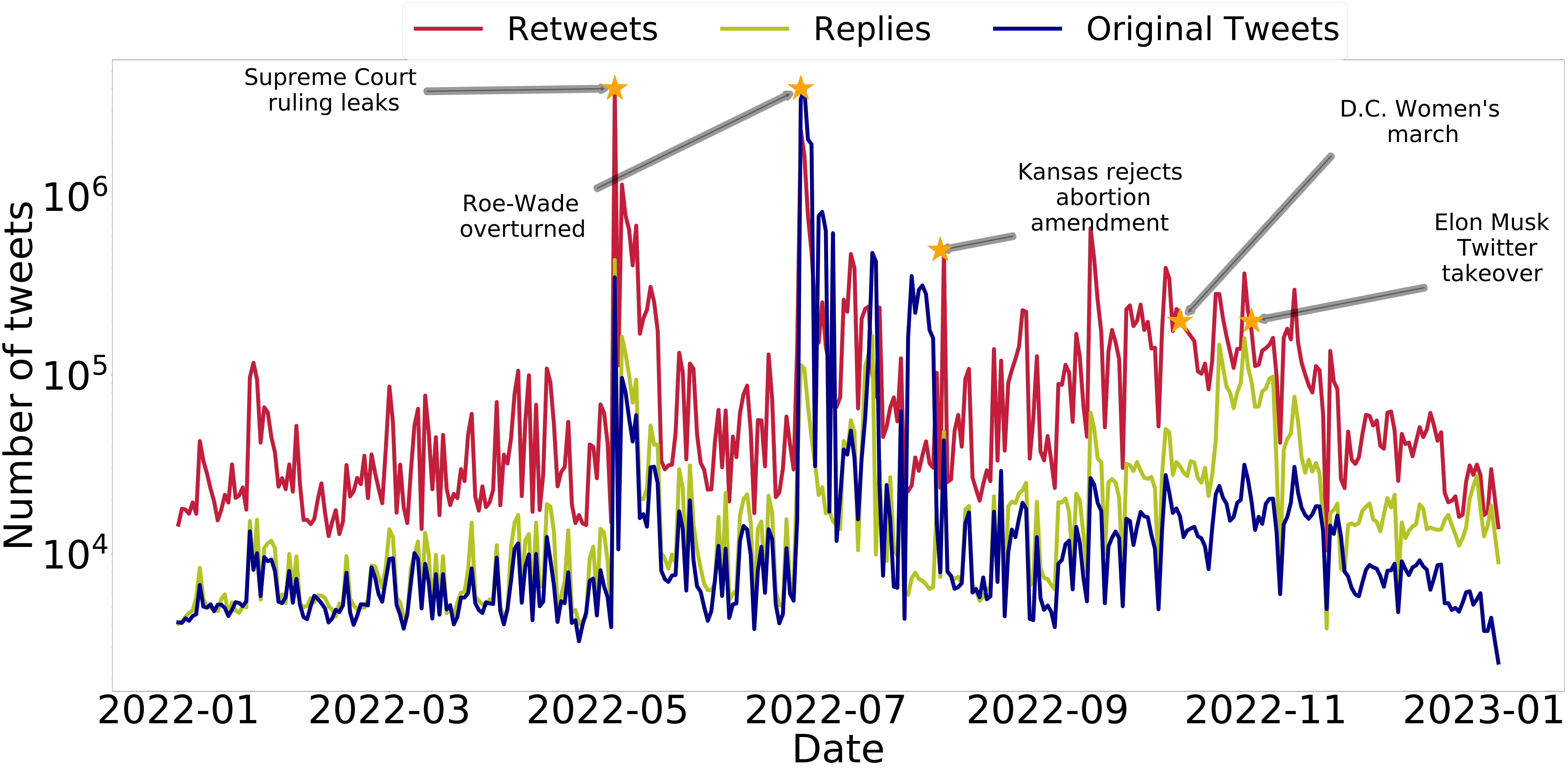}}
\caption{Daily Tweet Activity: Number of original tweets, retweets, and replies over time. Notable events are highlighted in the timeline}
\label{fig:tweet_counts}
\end{figure}

\section*{Descriptive Analysis}

\subsection*{Tweet Statistics}

\begin{table}[h]
\begin{tabularx}{\columnwidth}{|c|c|c|c|}
\toprule
\textbf{Tweets} & \textbf{Retweets} & \textbf{Replies} & \textbf{Quoted Tweets}\\ \hline
21,688,663 & 42,332,378 & 8,328,364 & 1,131,301\\\hline
\end{tabularx}
\caption{Tweet Statistics}
\label{tab:tweet_stats}
\end{table}

We identify original tweets, retweets, quote tweets, and replies in our dataset. If the tweet was a retweet, quote tweet, or a reply, the tweet object lists the ID of the user referenced in the interaction through ``retweeted user id'', ``quoted user id'' and ``in reply to user id'' fields respectively. The number of tweets, retweets, replies, and quote tweets in the dataset are shown in Table \ref{tab:tweet_stats}. 
In Fig \ref{fig:tweet_counts}, we show the time series of the total number of original tweets, retweets, and replies. Since quote tweets are low in number, they are not depicted in Fig \ref{fig:tweet_counts} to reduce clutter. We also highlight crucial events relating to Roe v. Wade as vertical lines. The absolute volume of tweets using our keywords has an abrupt increase on May 2nd, 2022, when a draft decision of the Supreme Court was leaked to the press. 
Similarly, we also see a spike in tweets on June 24th, 2022 when the Supreme Court officially overturned its 1973 ruling in Roe v. Wade (Refer Fig \ref{fig:tweet_counts}. During the midterm elections in November 2022, we also see a spike in reply counts, hinting at an increased discussion about abortion rights.  

\begin{figure*}
    \subfigure[Top-20 hashtags shared]
    {\includegraphics[width=0.49\linewidth]{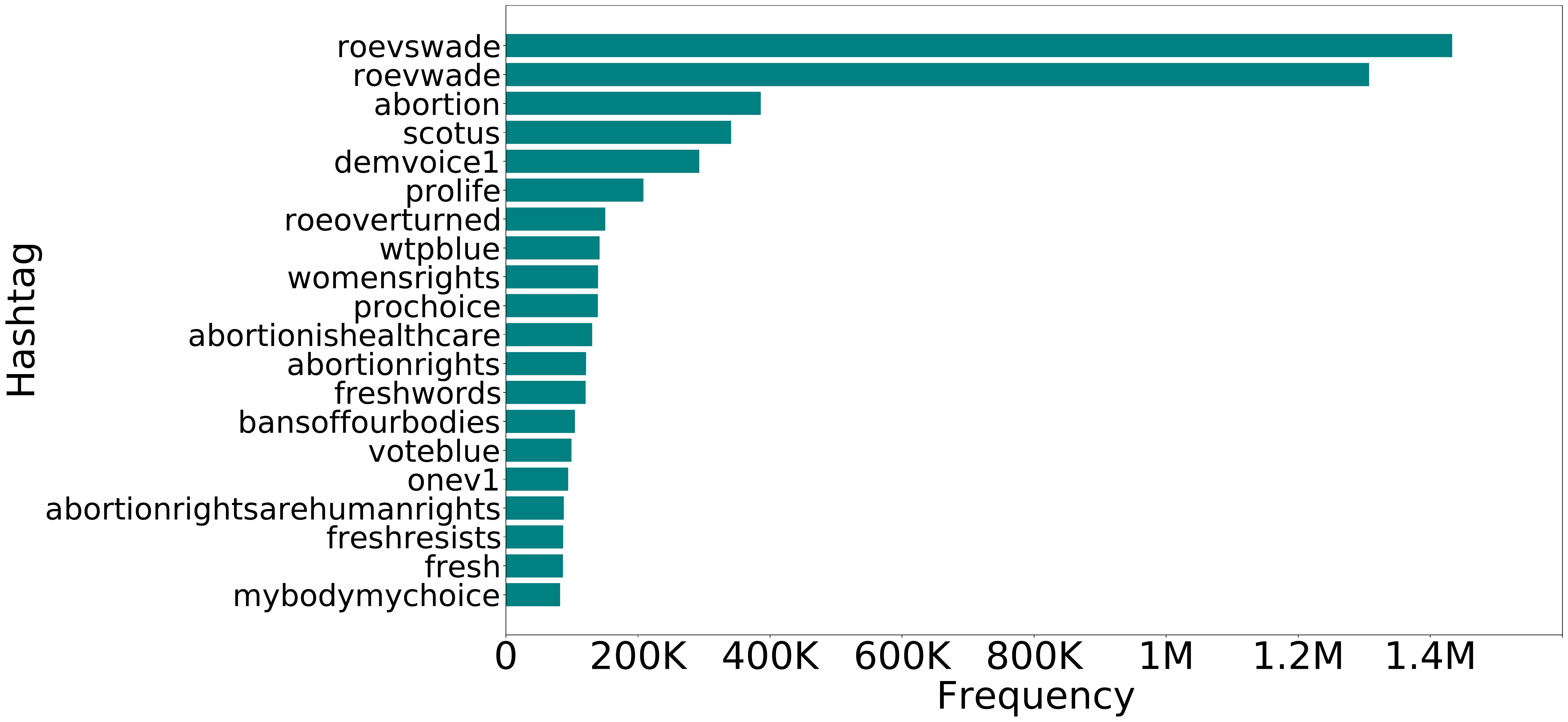}}
    \subfigure[Hashtag use over time]
    {\includegraphics[width=0.49\linewidth]{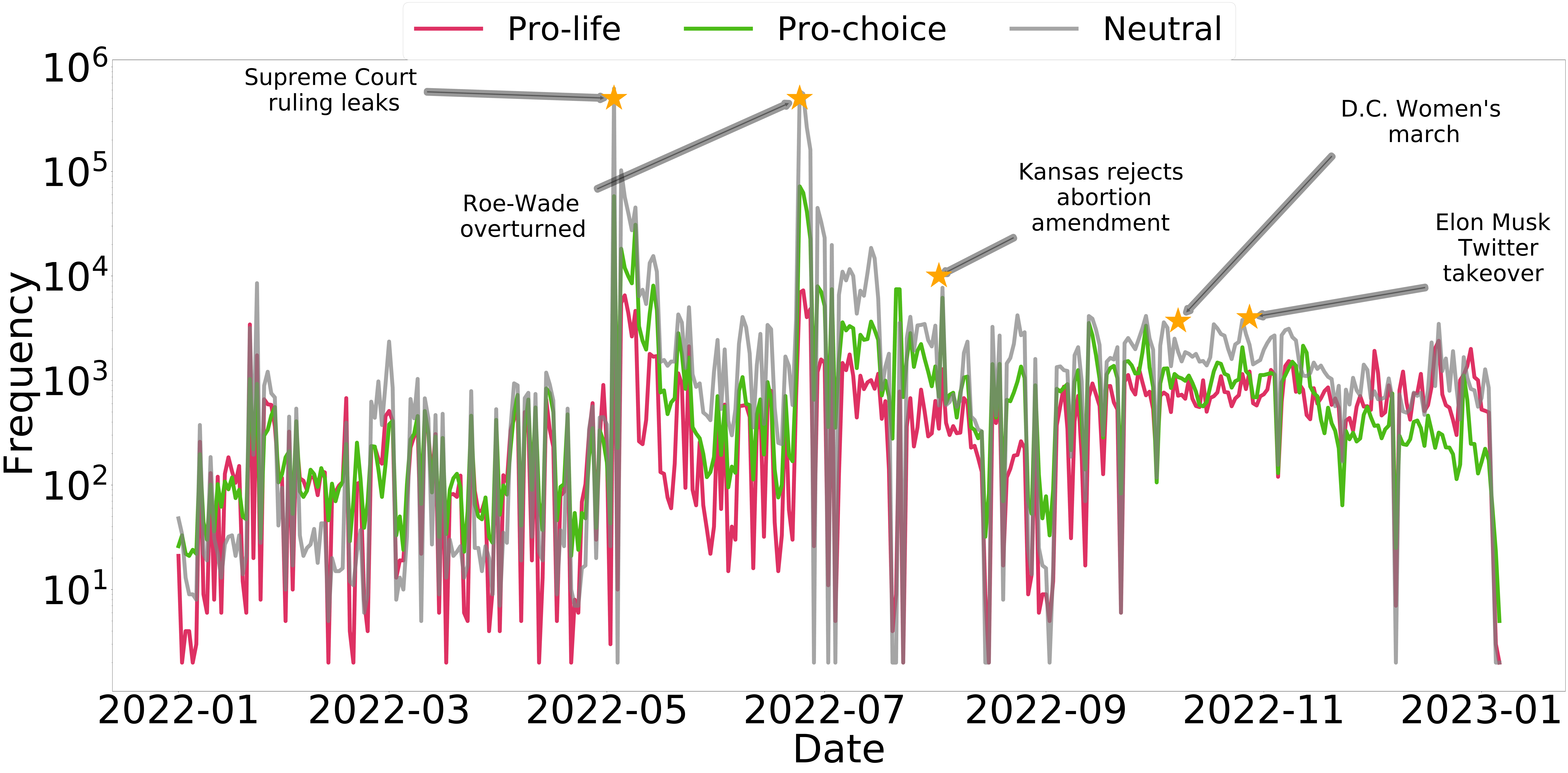}}
\caption{(a) Top-20 hashtags shared by users in our dataset. (b) Usage of Pro-life, Pro-Choice and Neutral hashtags over time, including hashtags that co-occur with \#prolife, \#prochoice and \#roevwade topics respectively. Refer to hashtags in Table \ref{tab:hash}. Notable events are highlighted in the timeline.}
\label{fig:hash_counts}
\end{figure*}

\begin{figure}
    {\includegraphics[width=0.98\linewidth]{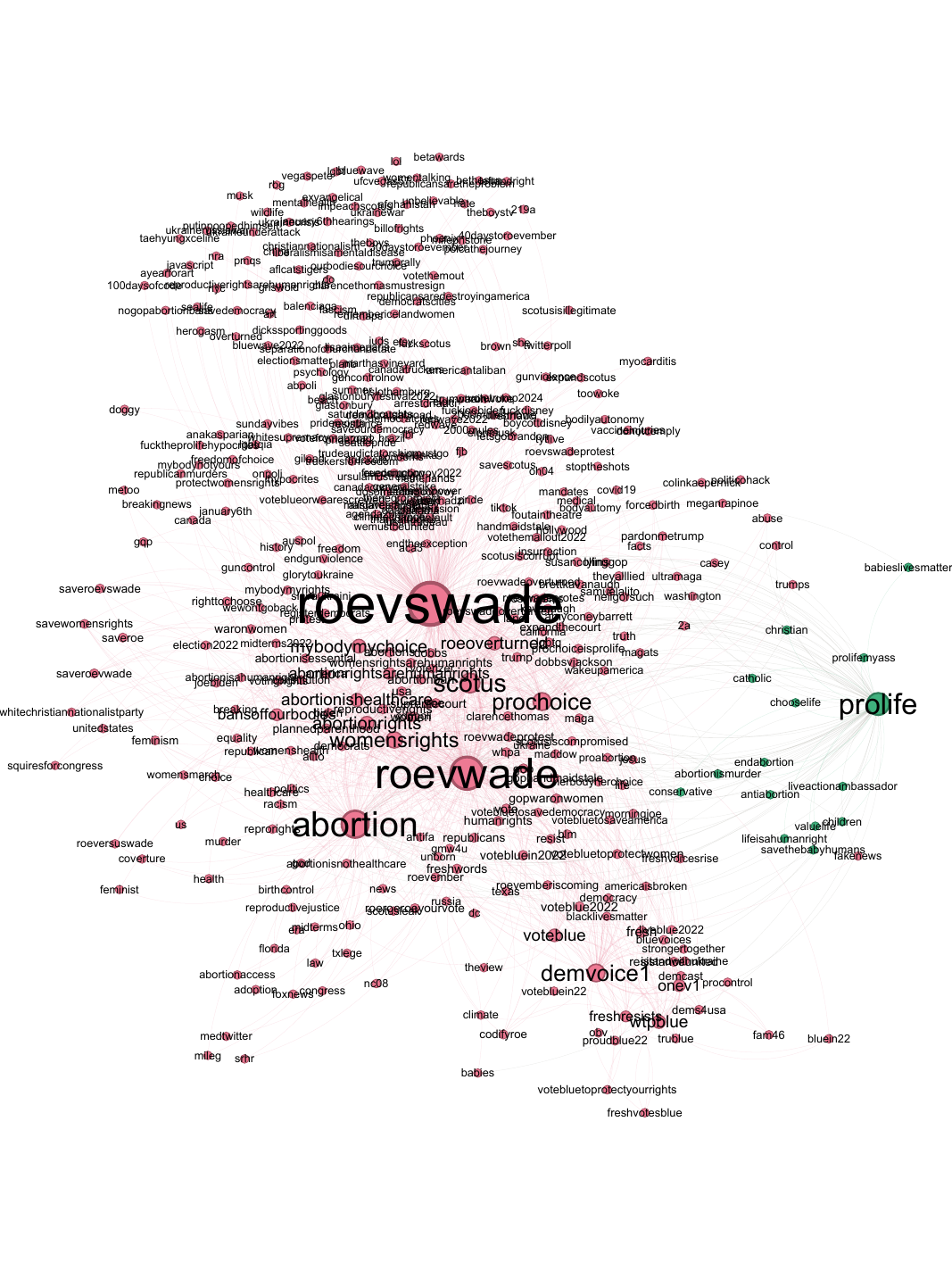}}
\caption{Clustering in hashtag co-occurrence network. The pink cluster shows prominent pro-choice hashtags whereas, the green cluster highlights pro-life hashtags.}
\label{fig:hash_occur}
\end{figure}

\subsection*{Hashtags}

Each tweet object consists of an entities field, which lists the hashtags used in the tweet. In Fig \ref{fig:hash_counts}(a), we show that among the top 20 tweeted hashtags, most are directly related to the abortion rights debate in the US. 
The prominent usage of hashtags like \#roevswade, \#prolife, \#roeoverturned, \#prochoice, \#abortionishealthcare, \#freshwords, and \#mybodymychoice reflect both sides of the debate. 
We also show the usage of pro-choice, pro-life, and neutral hashtags (Refer Table \ref{tab:hash}) in Fig \ref{fig:hash_counts}(b). We plot this time series using the hashtags in our keyword set in Table \ref{tab:hash}. Along expected lines, neutral hashtags like are used the most. These hashtags can co-occur prominently with both sides of the abortion rights debate. Pro-choice hashtags largely remain more prominent than pro-life hashtags, with both groups seeing significant spikes around May 2nd, 2022 and June 24th, 2022.  

Generally, hashtags are used to highlight the content of tweets. In some cases, users prefer to use multiple hashtags related to one another. We identify pairs of hashtags that appear together in tweets to build a hashtag co-occurrence network. To isolate related hashtags, we cluster them by applying the Label Propagation community detection algorithm \cite{raghavan2007near}. Python-igraph's implementation of the algorithm \cite{igraph2006} allows us to specify a seed set of hashtags which we can specify have  the community (pro-life/pro-choice) they belong to based on our domain-specific knowledge. We use the set of 28 pro-choice and pro-life hashtags in Table \ref{tab:hash} as the seed set. Given the small numbers, we avoid including neutral hashtags in the seed set. 

For better visualization, we only consider edges that connect at least one seed node. We also ignore infrequently co-occurring hashtags by setting a threshold of 500 on minimum co-occurrences. Clusters obtained through Label Propagation are shown in Fig \ref{fig:hash_occur}. The pink cluster shows prominent pro-choice hashtags, and the green cluster highlights the pro-life hashtags.  Hashtags such as \#prochoiceisprolife, \#saveroevwade frequently co-occur with pro-choice hashtags while, \#children and \#babieslivesmatter appear with pro-life hashtags. Given that neutral hashtags such as \#roevswade and \#scotus co-occurr more frequently with pro-choice hashtags, they get clustered into the pro-choice group. While Fig \ref{fig:hash_occur} shows that prochoice hashtags seemingly dwarf pro-life hashtags, it is worth noting that in Fig \ref{fig:hash_counts}(a), \#prolife appears more commonly than \#prochoice in our dataset. The smaller pro-life cluster could be because other pro-life hashtags aren't quite as popular, whereas several other pro-choice hashtags are prominent (Refer Fig \ref{fig:hash_counts}(a)). 
Additionally, Twitter being more popular among liberals \cite{pew} could potentially be another factor.

%
\begin{figure*}
\subfigure[Top-20 domains shared]
{\includegraphics[width=0.49\linewidth]{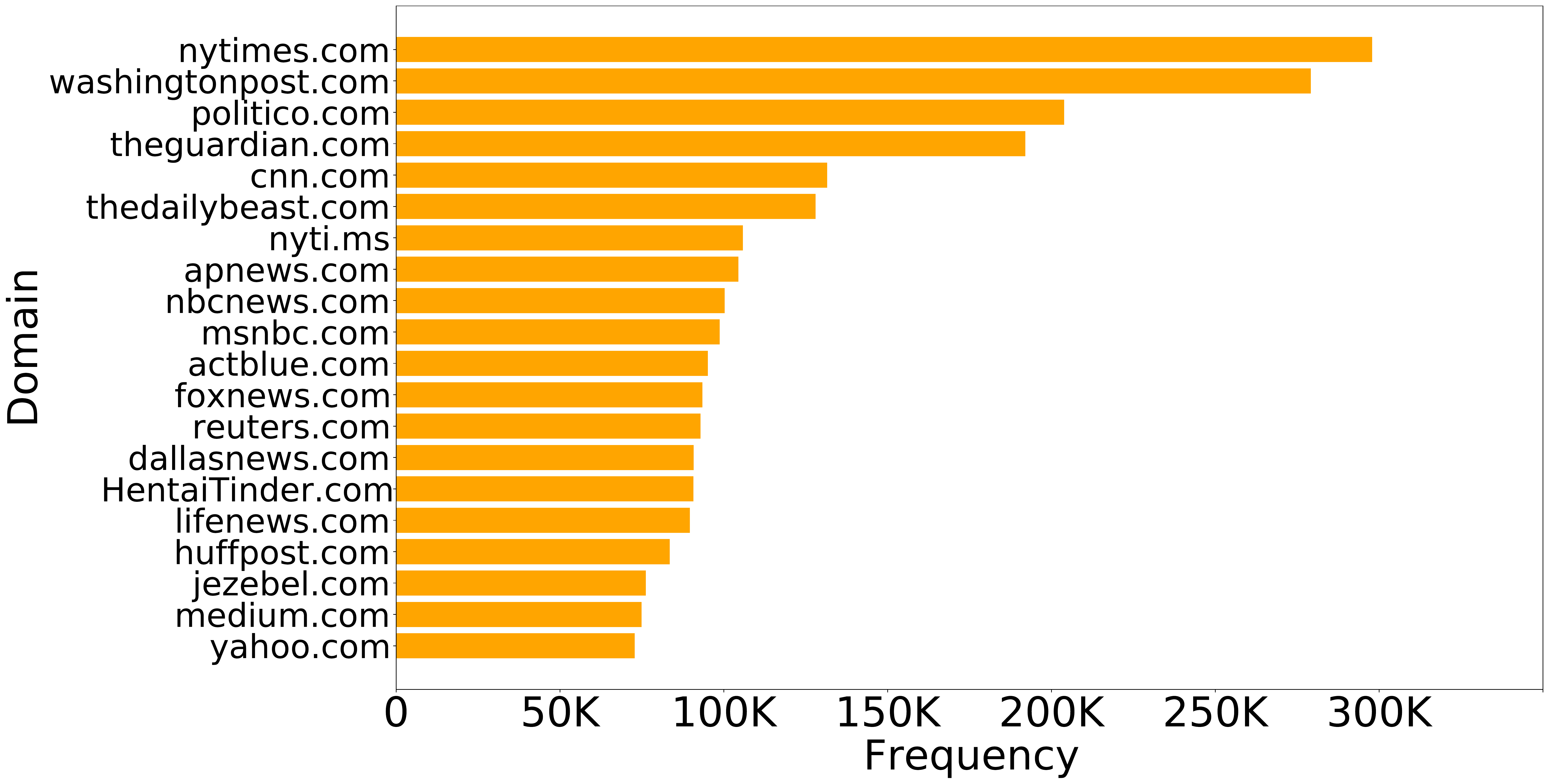}}
\subfigure[Domains: Ideology vs Reliability]
{\includegraphics[width=0.49\linewidth]{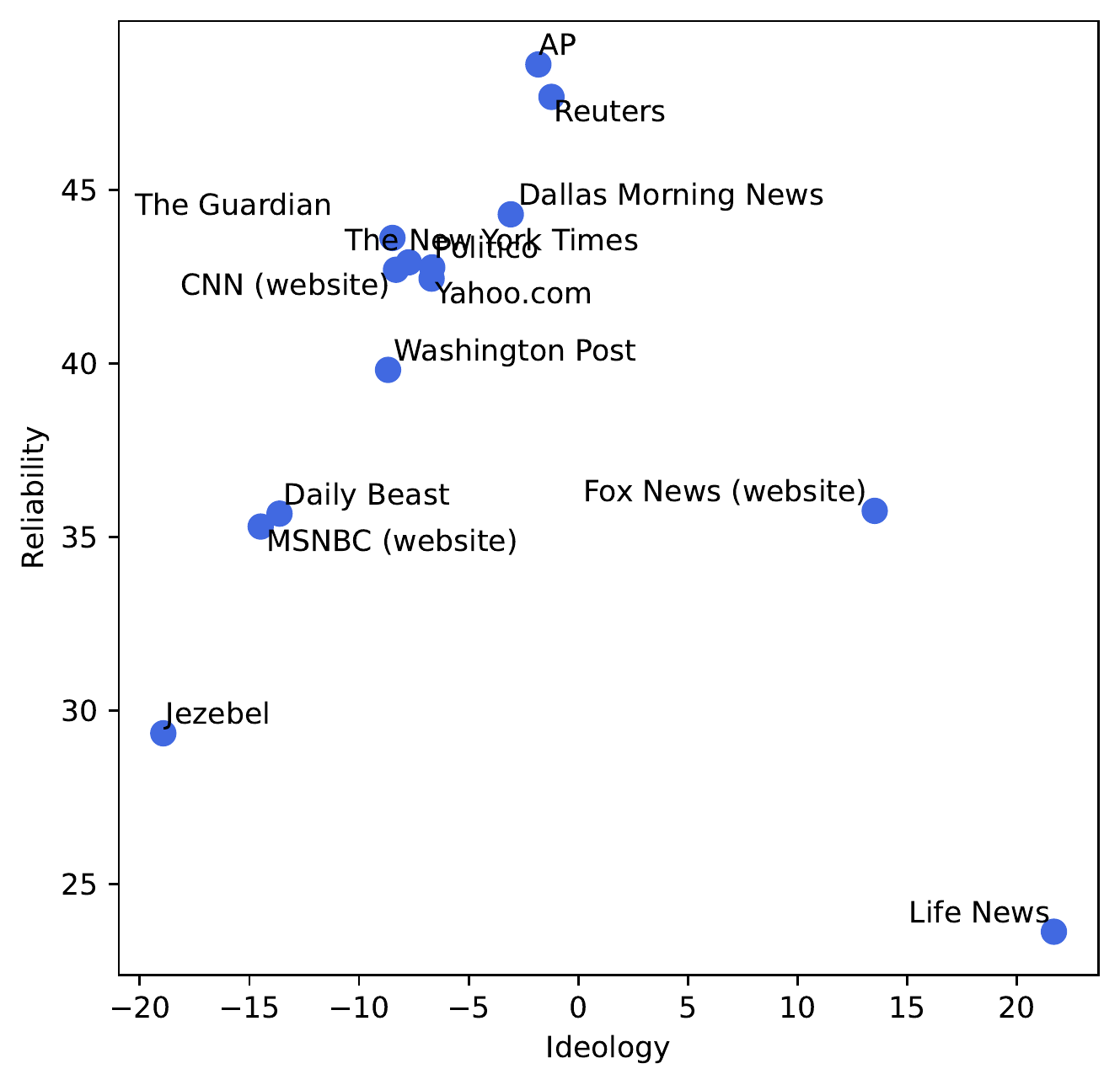}}
\caption{ (a) Top-20 domains shared by users. (b) Top-20 domains with ideology and reliability scores from Media Bias - Ad Fontes.}
\label{fig:domain_counts}
\end{figure*}

\subsection*{Domains}


The entities field in tweet objects includes lists of the URL embedded in the tweet, if any. We use \textit{tldextract}\footnote{\url{https://pypi.org/project/tldextract/}} Python package to extract the pay-level domain of the URLs shared. In Fig \ref{fig:domain_counts}(a), we show the top 20 most commonly shared pay-level domains. Most of these domains are prominent media sources and consequently have their political leaning.  
We were able to identify 14 of the top 20 domains through Ad Fontes Media-Bias \cite{ad2020media} \footnote{Ad Fontes Media-Bias Chart is a two-dimensional chart that positions 100 media outlets in terms of their political ideologies and content qualities. Ad Fontes evaluates media outlets based on ratings from 1,818 online articles and 98 cable news shows.\cite{heldebrandt2019popular}}. 
The Ad Fontes Media-Bias provides ideology score and content reliability score\footnote{On the Ad Fontes ideology-slant classification, the ideology score ranges from -42 to 42, representing extreme left to extreme right political leaning. The 0 in ideology score represents moderate or centrist political leaning.} for domains. 
Among the 14 domains in our dataset in Fig \ref{fig:domain_counts}(b), 12 were left-leaning domains, with an overall mean of $ -4.61$. In addition, the majority (i.e., 12) of our top domains have a reliability score higher than 30, with the reliability or news value score ranging from 0 to 64, representing low to high reliability. 


\subsection*{Retweet Network}

Given that we have over $44M$ retweet interactions, we can extract a retweet network of users. From each of these $44M$ retweet interactions, we extract user IDs of both the account making the retweet (user id) and the account being retweeted (retweeted user id) and create a retweet network with $6.1M$ users and $27M$ unique interactions between them. Previous studies have leveraged retweet networks to not only study its properties \cite{bild2015aggregate} but also explore polarization and exposures \cite{conover2011political,rao2022partisan}. 

\begin{table}[hb]
\centering
\begin{tabularx}{6cm}{p{3cm}| p{3cm}}
\toprule
\textbf{Statistic} & \textbf{Value}\\ \hline
Nodes & 6,089,939\\\hline
Edges & 26,925,519\\\hline
Max. Indegree & 196,085 \\\hline
Max. Outdegree & 9,558\\\hline
Density & 7.2 x $10^{-7}$ \\\hline
\end{tabularx}
\caption{RT Network Statistics}
\label{tab:net_stats}
\end{table}

\begin{figure}
    {\includegraphics[width=0.98\linewidth]{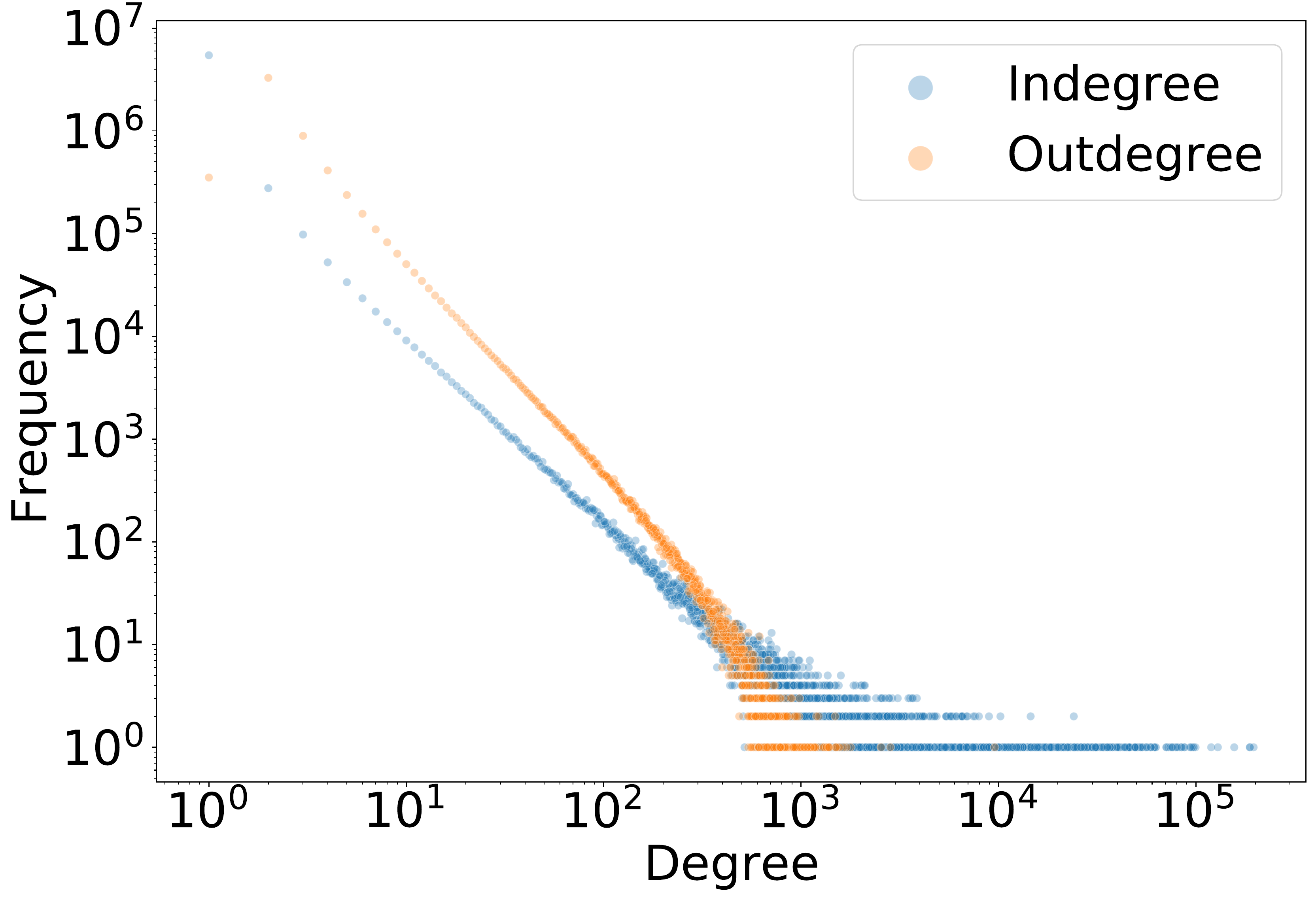}}
\caption{Indegree and Outdegree distribution in the Retweet Network}
\label{fig:deg_dist}
\end{figure}

In Table \ref{tab:net_stats}, we list simple statistics to summarize the retweet network. The most retweeted user has been retweeted for over $196K$ times and the user with the highest number of retweets has $9.5K$ retweets. Retweet interactions are usually not reciprocal. A small fraction of users gets retweeted most of the time. This is reflected by the network density statistic in Table \ref{tab:net_stats}. The log-log plot of degree distributions in Fig \ref{fig:deg_dist} shows that the in-degree and out-degree distributions of the retweet network follow a power law.

\section*{Release and Access}
Our dataset is publicly available for download on Harvard Dataverse. \footnote{\url{https://dataverse.harvard.edu/dataset.xhtml?persistentId=doi:10.7910/DVN/STU0J5}}. The repository consists of Comma-Separated Value (CSV) files of tweet IDs separated by date. In order to extract tweet objects from Twitter IDs, users of this dataset can rely on third-party re-hydration tools - Hydrator \cite{hydrator} or the Twitter API's search endpoint \cite{twitter2022search}.  

\section*{Discussion}
%


On May 2nd, 2022, a leaked draft decision of the Supreme Court in Dobbs v. Jackson Women's Health Organization was published by Politico \cite{leak2022politico}. In this initial draft majority opinion, Supreme Court justices had voted and opined in favor of overruling Roe v. Wade. This triggered a fierce outcry from pro-choice individuals and groups. Social media platforms became sites for contentious debate yet again. Pro-choice protests and mobilization efforts were met with equal measure from pro-life activists and groups. On June 24th, 2022, the Supreme Court officially ruled to set aside rulings made in Roe v. Wade, thereby paving way for certain states to outlaw abortion.

In light of these events have triggered increased discussions, many of which occurred on social media platforms.
It is crucial to preserve the traces of these discussions in order to allow researchers to study and understand the implications of the ruling on online public opinion.  
Toward this end, we release the first large-scale Twitter dataset comprising of over $ 73$ million in tweets identified systematically with keywords from both sides of the contentious debate: pro-abortion and anti-abortion, over the course of one year.  
The broad coverage of our dataset makes the study of a wide variety of research problems possible. 
The dataset has the potential and can benefit the research community in several ways, we listed a few possible usage below. 

\subsubsection{Opinion Dynamics and Polarization}
Our dataset can be used to study opinion dynamics and the spread of pro-life and pro-choice communities online. 
For example, comparing the number of retweets, the speed of the diffusion, retweet network structure, and how different ego characters such as politicians or celebrities plays different roles between pro-life and pro-choice communities.  
Given that we have a clear stance on the topic of abortion, polarization and the development of echo chambers between pro-life and pro-choice communities can be intriguing avenues and could be studied using our dataset.  

\subsubsection{Protest Mobilization}

While the role of social media in protest mobilization has been studied for specific wedge issues or events \cite{ince2017social, wolfsfeld2013social, steinert2017spontaneous, freelon2018quantifying}, studies comparing interactions around different protests are sparse \cite{segerberg2011social,won2017protest}. This dataset can be used in conjunction with other datasets to study complex social movements such as the Black Live Matter movement \cite{giorgi2022twitter}, and capital riots \cite{dan2021capitol}. 
For example, do these social movements rely on certain agents' call for action? Do discussion networks' size and density correlate with the public opinions on the protests or the media's coverage of the protests? If we could quantify the magnitude of each offline protest, could one predict it using online data? And what are the crucial components in the online discussions that determine the success or failure of an offline protest? 
Furthermore, this dataset facilitates the examination of how certain public figures, politicians, pundits, and also non-political actors, disseminate, support, or engage with different claims related to Roe v. Wade and abortion more broadly. One could also systematically examine the relative contribution of political elites (i.e., politicians, journalists, pundits, and news media organizations) versus the “ordinary users” to the overall volume and nature of these online discussions. Did ordinary citizens contribute more abortion-related tweets than politicians? How did these patterns vary by political leaning of the citizens and the politicians?

\subsubsection{Emotion, Moral Attitudes, and Multi-Modalities}
Different groups express different moral attitudes and emotions to specific events. Multi-modal information in the form of text, images, embedded videos, and hyperlinks can be extracted and analyzed to quantify them. Additionally, one can attempt to quantify agreeableness and sarcasm. Some possible research questions include - whether pro-choice or pro-life individuals express more anger or other negative sentiments, whether certain political parties use more propaganda techniques in abortion-related messages, and whether it correlated with the midterm election outcome. Furthermore, one could ask how hate speech toward woman stigmatize abortion and plays a role in abortion-related debates.

\subsubsection{Bots and Misinformation}
The rise of social bots has paved ways for malicious actors to infiltrate political discourse and spread misinformation \cite{ferrara2016rise}. Exploring the influence of bot accounts in abortion rights-related discourse and the subsequent harms posed by health-related misinformation is another interesting area of research. Potential research questions include whether there was an increase in health-related misinformation once Roe v. Wade was overturned, what agenda was propagated through bots, and whether exposure to such discourse can influence individual opinions.   




Overall, this dataset provides an important source to study the dynamics of online discourse around abortion rights in the US. In future work, we will use computational methods to quantify the stances on abortion rights in the tweets and measure the level of toxicity expressed in those stances. 

\section*{Limitations}

Our data collection relied on a list of relevant keywords, which may not fully capture the extent of the event. Additionally, the dataset only includes English-language tweets, and may not capture multilingual discourse relating to the topic. The data collection also relies on both the stream and archive search endpoints of the Twitter API, and while the data covers every day in 2022, it may not include all tweets if they were deleted before collection. Other potential factors that may impact the data collection include changes in ownership at Twitter and subsequent changes to the Trust and Safety team, which could affect the API's functionality. Additionally, the dataset only includes data from Twitter, which has been shown to be more popular among liberals \cite{pew}. Despite these limitations, the keywords were chosen systematically and the dataset's validity has been demonstrated in its statistics.

\section*{Broader perspective and ethics}

All of the data in this dataset is publicly available information. Our data collection was deemed exempt from review by the Institutional Review Board (IRB) at the University of California-Davis and University of Southern California, as it relied solely on publicly available data. The data collection adheres to Twitter's terms of service \cite{twitter2022policy}. In compliance with Twitter's terms and conditions, we only release the Tweet IDs of publicly available tweets and require users of the dataset to comply with Twitter's terms and conditions as well.

There may be ethical concerns regarding user anonymity. Tweet objects contain user information. Users have the option to restrict their tweets from being made available through the API by switching to a private account or by deleting their tweets. In this article, we only present aggregated statistics to address this concern.

\bibliography{ref}

\begin{thebibliography}{56}
\providecommand{\natexlab}[1]{#1}

\bibitem[{Abilov et~al.(2021)Abilov, Hua, Matatov, Amir, and
  Naaman}]{abilov2021voterfraud2020}
Abilov, A.; Hua, Y.; Matatov, H.; Amir, O.; and Naaman, M. 2021.
\newblock Voterfraud2020: a multi-modal dataset of election fraud claims on
  Twitter.
\newblock \emph{arXiv preprint arXiv:2101.08210}.

\bibitem[{Abramowitz and Saunders(2008)}]{abramowitz2008polarization}
Abramowitz, A.~I.; and Saunders, K.~L. 2008.
\newblock Is polarization a myth?
\newblock \emph{The Journal of Politics}, 70(2): 542--555.

\bibitem[{Barber{\'a}(2015)}]{barbera2015birds}
Barber{\'a}, P. 2015.
\newblock Birds of the same feather tweet together: Bayesian ideal point
  estimation using Twitter data.
\newblock \emph{Political analysis}, 23(1): 76--91.

\bibitem[{Barber{\'a} et~al.(2019)Barber{\'a}, Casas, Nagler, Egan, Bonneau,
  Jost, and Tucker}]{barbera2019leads}
Barber{\'a}, P.; Casas, A.; Nagler, J.; Egan, P.~J.; Bonneau, R.; Jost, J.~T.;
  and Tucker, J.~A. 2019.
\newblock Who leads? Who follows? Measuring issue attention and agenda setting
  by legislators and the mass public using social media data.
\newblock \emph{American Political Science Review}, 113(4): 883--901.

\bibitem[{BESTVATER et~al.(2022)BESTVATER, SHAH, RIVER, and
  SMITH}]{bestvater2022politics}
BESTVATER, S.; SHAH, S.; RIVER, G.; and SMITH, A. 2022.
\newblock Politics on twitter: One-third of tweets from us adults are
  political.

\bibitem[{Bild et~al.(2015)Bild, Liu, Dick, Mao, and
  Wallach}]{bild2015aggregate}
Bild, D.~R.; Liu, Y.; Dick, R.~P.; Mao, Z.~M.; and Wallach, D.~S. 2015.
\newblock Aggregate characterization of user behavior in Twitter and analysis
  of the retweet graph.
\newblock \emph{ACM Transactions on Internet Technology (TOIT)}, 15(1): 1--24.

\bibitem[{Brenan(2010)}]{gallup2021abortion}
Brenan, M. 2010.
\newblock Record-High 47\% in U.S. Think Abortion Is Morally Acceptable.
\newblock
  \url{https://news.gallup.com/poll/350756/record-high-think-abortion-morally-acceptable.aspx}.
\newblock [Online; accessed 11-January-2023].

\bibitem[{Breuer, Landman, and Farquhar(2015)}]{breuer2015social}
Breuer, A.; Landman, T.; and Farquhar, D. 2015.
\newblock Social media and protest mobilization: Evidence from the Tunisian
  revolution.
\newblock \emph{Democratization}, 22(4): 764--792.

\bibitem[{Chen et~al.(2021)Chen, Chang, Rao, Lerman, Cowan, and
  Ferrara}]{chen2021covid}
Chen, E.; Chang, H.; Rao, A.; Lerman, K.; Cowan, G.; and Ferrara, E. 2021.
\newblock COVID-19 misinformation and the 2020 US presidential election.
\newblock \emph{The Harvard Kennedy School Misinformation Review}.

\bibitem[{Chen, Deb, and Ferrara(2022)}]{chen2022election2020}
Chen, E.; Deb, A.; and Ferrara, E. 2022.
\newblock \# Election2020: The first public Twitter dataset on the 2020 US
  Presidential election.
\newblock \emph{Journal of Computational Social Science}, 5(1): 1--18.

\bibitem[{Chen and Ferrara(2022)}]{chen2022tweets}
Chen, E.; and Ferrara, E. 2022.
\newblock Tweets in time of conflict: A public dataset tracking the twitter
  discourse on the war between ukraine and russia.
\newblock \emph{arXiv preprint arXiv:2203.07488}.

\bibitem[{Chen et~al.(2020)Chen, Lerman, Ferrara et~al.}]{chen2020tracking}
Chen, E.; Lerman, K.; Ferrara, E.; et~al. 2020.
\newblock Tracking social media discourse about the covid-19 pandemic:
  Development of a public coronavirus twitter data set.
\newblock \emph{JMIR public health and surveillance}, 6(2): e19273.

\bibitem[{Cinelli et~al.(2021)Cinelli, De~Francisci~Morales, Galeazzi,
  Quattrociocchi, and Starnini}]{cinelli2021echo}
Cinelli, M.; De~Francisci~Morales, G.; Galeazzi, A.; Quattrociocchi, W.; and
  Starnini, M. 2021.
\newblock The echo chamber effect on social media.
\newblock \emph{Proceedings of the National Academy of Sciences}, 118(9):
  e2023301118.

\bibitem[{Conover et~al.(2011)Conover, Ratkiewicz, Francisco, Gon{\c{c}}alves,
  Menczer, and Flammini}]{conover2011political}
Conover, M.; Ratkiewicz, J.; Francisco, M.; Gon{\c{c}}alves, B.; Menczer, F.;
  and Flammini, A. 2011.
\newblock Political polarization on twitter.
\newblock In \emph{Proceedings of the international aaai conference on web and
  social media}, volume~5, 89--96.

\bibitem[{Csardi and Nepusz(2006)}]{igraph2006}
Csardi, G.; and Nepusz, T. 2006.
\newblock The igraph software package for complex network research.
\newblock \emph{InterJournal}, Complex Systems: 1695.

\bibitem[{DiMaggio, Evans, and Bryson(1996)}]{dimaggio1996have}
DiMaggio, P.; Evans, J.; and Bryson, B. 1996.
\newblock Have American's social attitudes become more polarized?
\newblock \emph{American journal of Sociology}, 102(3): 690--755.

\bibitem[{DocNow(2020)}]{hydrator}
DocNow. 2020.
\newblock Hydrator [Computer Software].
\newblock \url{https://github.com/docnow/hydrator}.
\newblock [Online; accessed 11-January-2023].

\bibitem[{Evans(2003)}]{evans2003have}
Evans, J.~H. 2003.
\newblock Have Americans' attitudes become more polarized?—An update.
\newblock \emph{Social Science Quarterly}, 84(1): 71--90.

\bibitem[{Ferrara et~al.(2016)Ferrara, Varol, Davis, Menczer, and
  Flammini}]{ferrara2016rise}
Ferrara, E.; Varol, O.; Davis, C.; Menczer, F.; and Flammini, A. 2016.
\newblock The rise of social bots.
\newblock \emph{Communications of the ACM}, 59(7): 96--104.

\bibitem[{Fiorina, Abrams et~al.(2008)}]{fiorina2008political}
Fiorina, M.~P.; Abrams, S.~J.; et~al. 2008.
\newblock Political polarization in the American public.
\newblock \emph{ANNUAL REVIEW OF POLITICAL SCIENCE-PALO ALTO-}, 11: 563.

\bibitem[{Founta et~al.(2018)Founta, Djouvas, Chatzakou, Leontiadis, Blackburn,
  Stringhini, Vakali, Sirivianos, and Kourtellis}]{founta2018large}
Founta, A.~M.; Djouvas, C.; Chatzakou, D.; Leontiadis, I.; Blackburn, J.;
  Stringhini, G.; Vakali, A.; Sirivianos, M.; and Kourtellis, N. 2018.
\newblock Large scale crowdsourcing and characterization of twitter abusive
  behavior.
\newblock In \emph{Twelfth International AAAI Conference on Web and Social
  Media}.

\bibitem[{Freelon, McIlwain, and Clark(2018)}]{freelon2018quantifying}
Freelon, D.; McIlwain, C.; and Clark, M. 2018.
\newblock Quantifying the power and consequences of social media protest.
\newblock \emph{New Media \& Society}, 20(3): 990--1011.

\bibitem[{Garimella et~al.(2018)Garimella, De~Francisci~Morales, Gionis, and
  Mathioudakis}]{garimella2018political}
Garimella, K.; De~Francisci~Morales, G.; Gionis, A.; and Mathioudakis, M. 2018.
\newblock Political discourse on social media: Echo chambers, gatekeepers, and
  the price of bipartisanship.
\newblock In \emph{Proceedings of the 2018 world wide web conference},
  913--922.

\bibitem[{Gerstein and Ward(2020)}]{leak2022politico}
Gerstein, J.; and Ward, A. 2020.
\newblock Supreme Court has voted to overturn abortion rights, draft opinion
  shows.
\newblock
  \url{https://www.politico.com/news/2022/05/02/supreme-court-abortion-draft-opinion-00029473}.

\bibitem[{Giorgi et~al.(2022)Giorgi, Guntuku, Himelein-Wachowiak, Kwarteng,
  Hwang, Rahman, and Curtis}]{giorgi2022twitter}
Giorgi, S.; Guntuku, S.~C.; Himelein-Wachowiak, M.; Kwarteng, A.; Hwang, S.;
  Rahman, M.; and Curtis, B. 2022.
\newblock Twitter Corpus of the\# BlackLivesMatter Movement and Counter
  Protests: 2013 to 2021.
\newblock In \emph{Proceedings of the International AAAI Conference on Web and
  Social Media}, volume~16, 1228--1235.

\bibitem[{Guo et~al.(2022)Guo, Burghardt, Rao, and Lerman}]{guo2022emotion}
Guo, S.; Burghardt, K.; Rao, A.; and Lerman, K. 2022.
\newblock Emotion Regulation and Dynamics of Moral Concerns During the Early
  COVID-19 Pandemic.
\newblock \emph{arXiv preprint arXiv:2203.03608}.

\bibitem[{Heldebrandt(2019)}]{heldebrandt2019popular}
Heldebrandt, B. 2019.
\newblock How a popular media bias chart determines what news can be trusted.
\newblock \emph{Gateway Journalism Review}, 48(355): 20--22.

\bibitem[{Huang et~al.(2020)Huang, Jamison, Broniatowski, Quinn, and
  Dredze}]{huang_xiaolei_2020_3735015}
Huang, X.; Jamison, A.; Broniatowski, D.; Quinn, S.; and Dredze, M. 2020.
\newblock Coronavirus Twitter Data: A collection of COVID-19 tweets with
  automated annotations.
\newblock Http://twitterdata.covid19dataresources.org/index.

\bibitem[{Ince, Rojas, and Davis(2017)}]{ince2017social}
Ince, J.; Rojas, F.; and Davis, C.~A. 2017.
\newblock The social media response to Black Lives Matter: How Twitter users
  interact with Black Lives Matter through hashtag use.
\newblock \emph{Ethnic and racial studies}, 40(11): 1814--1830.

\bibitem[{Kerchner and Wrubel(2021)}]{dan2021capitol}
Kerchner, D.; and Wrubel, L. 2021.
\newblock {U.S. Capitol Riot and \#TrumpRally Tweet IDs}.

\bibitem[{McGregor(2019)}]{mcgregor2019social}
McGregor, S.~C. 2019.
\newblock Social media as public opinion: How journalists use social media to
  represent public opinion.
\newblock \emph{Journalism}, 20(8): 1070--1086.

\bibitem[{Media(2020)}]{ad2020media}
Media, A.~F. 2020.
\newblock The media bias chart.

\bibitem[{Mouw and Sobel(2001)}]{mouw2001culture}
Mouw, T.; and Sobel, M.~E. 2001.
\newblock Culture wars and opinion polarization: the case of abortion.
\newblock \emph{American Journal of Sociology}, 106(4): 913--943.

\bibitem[{Munn(2021)}]{munn2021more}
Munn, L. 2021.
\newblock More than a mob: Parler as preparatory media for the US Capitol
  storming.
\newblock \emph{First Monday}.

\bibitem[{NAF(2022)}]{naf2022abortion}
NAF. 2022.
\newblock 2021 Violence and Disruption Report.
\newblock
  \url{https://prochoice.org/national-abortion-federation-releases-2021-violence-disruption-report/}.
\newblock [Online; accessed 9-January-2023].

\bibitem[{Nelson and Tandoc~Jr(2019)}]{nelson2019doing}
Nelson, J.~L.; and Tandoc~Jr, E.~C. 2019.
\newblock Doing “well” or doing “good”: What audience analytics reveal
  about journalism’s competing goals.
\newblock \emph{Journalism Studies}, 20(13): 1960--1976.

\bibitem[{Nikolov, Flammini, and Menczer(2020)}]{nikolov2020right}
Nikolov, D.; Flammini, A.; and Menczer, F. 2020.
\newblock Right and left, partisanship predicts (asymmetric) vulnerability to
  misinformation.
\newblock \emph{arXiv preprint arXiv:2010.01462}.

\bibitem[{Pew(2022)}]{pew}
Pew. 2022.
\newblock Differences in How Democrats and Republicans Behave on Twitter.
\newblock
  \url{"https://www.pewresearch.org/politics/2020/10/15/differences-in-how-democrats-and-republicans-behave-on-twitter/"}.

\bibitem[{Priniski et~al.(2021)Priniski, Mokhberian, Harandizadeh, Morstatter,
  Lerman, Lu, and Brantingham}]{priniski2021mapping}
Priniski, J.~H.; Mokhberian, N.; Harandizadeh, B.; Morstatter, F.; Lerman, K.;
  Lu, H.; and Brantingham, P.~J. 2021.
\newblock Mapping moral valence of tweets following the killing of George
  Floyd.
\newblock \emph{arXiv preprint arXiv:2104.09578}.

\bibitem[{Raghavan, Albert, and Kumara(2007)}]{raghavan2007near}
Raghavan, U.~N.; Albert, R.; and Kumara, S. 2007.
\newblock Near linear time algorithm to detect community structures in
  large-scale networks.
\newblock \emph{Physical review E}, 76(3): 036106.

\bibitem[{Rao et~al.(2021)Rao, Morstatter, Hu, Chen, Burghardt, Ferrara, Lerman
  et~al.}]{rao2021political}
Rao, A.; Morstatter, F.; Hu, M.; Chen, E.; Burghardt, K.; Ferrara, E.; Lerman,
  K.; et~al. 2021.
\newblock Political partisanship and antiscience attitudes in online
  discussions about COVID-19: Twitter content analysis.
\newblock \emph{Journal of medical Internet research}, 23(6): e26692.

\bibitem[{Rao, Morstatter, and Lerman(2022)}]{rao2022partisan}
Rao, A.; Morstatter, F.; and Lerman, K. 2022.
\newblock Partisan Asymmetries in Exposure to Misinformation.
\newblock \emph{Scientific Reports}, 12.

\bibitem[{Saad(2010{\natexlab{a}})}]{gallup2011abortion}
Saad, L. 2010{\natexlab{a}}.
\newblock Americans Still Split Along Pro-Choice Pro-Life Lines.
\newblock
  \url{https://news.gallup.com/poll/147734/americans-split-along-pro-choice-pro-life-lines.aspx}.
\newblock [Online; accessed 11-January-2023].

\bibitem[{Saad(2010{\natexlab{b}})}]{gallup2010abortion}
Saad, L. 2010{\natexlab{b}}.
\newblock Republicans', Dems' Abortion Views Grow More Polarized.
\newblock
  \url{https://news.gallup.com/poll/126374/republicans-dems-abortion-views-grow-polarized.aspx}.
\newblock [Online; accessed 11-January-2023].

\bibitem[{SCOTUS(2022)}]{dobbs2022roe}
SCOTUS. 2022.
\newblock Dobbs, State Health Officer of the Mississippi Department of Health
  vs Jackson Women's Health Organization.
\newblock \url{https://www.supremecourt.gov/opinions/21pdf/19-1392_6j37.pdf}.
\newblock [Online; accessed 11-January-2023].

\bibitem[{Segerberg and Bennett(2011)}]{segerberg2011social}
Segerberg, A.; and Bennett, W.~L. 2011.
\newblock Social media and the organization of collective action: Using Twitter
  to explore the ecologies of two climate change protests.
\newblock \emph{The Communication Review}, 14(3): 197--215.

\bibitem[{Steinert-Threlkeld(2017)}]{steinert2017spontaneous}
Steinert-Threlkeld, Z.~C. 2017.
\newblock Spontaneous collective action: Peripheral mobilization during the
  Arab Spring.
\newblock \emph{American Political Science Review}, 111(2): 379--403.

\bibitem[{Stocking, Barthel, and Grieco(2018)}]{stocking2018sources}
Stocking, G.; Barthel, M.; and Grieco, E. 2018.
\newblock Sources shared on Twitter: A case study on immigration. Pew Research
  Center.

\bibitem[{Twitter(2022{\natexlab{a}})}]{twitter2022policy}
Twitter. 2022{\natexlab{a}}.
\newblock Developer Agreement and Policy.
\newblock
  \url{https://developer.twitter.com/en/developer-terms/agreement-and-policy}.
\newblock [Online; accessed 11-January-2023].

\bibitem[{Twitter(2022{\natexlab{b}})}]{twitter2022search}
Twitter. 2022{\natexlab{b}}.
\newblock Tweets Lookup Endpoint.
\newblock
  \url{https://developer.twitter.com/en/docs/twitter-api/tweets/lookup/api-reference/get-tweets-id}.
\newblock [Online; accessed 11-January-2023].

\bibitem[{Wojcieszak et~al.(2021)Wojcieszak, Casas, Yu, Nagler, and
  Tucker}]{wojcieszak2021echo}
Wojcieszak, M.; Casas, A.; Yu, X.; Nagler, J.; and Tucker, J.~A. 2021.
\newblock Echo chambers revisited: The (overwhelming) sharing of in-group
  politicians, pundits and media on Twitter.

\bibitem[{Wojcieszak et~al.(2022)Wojcieszak, Casas, Yu, Nagler, and
  Tucker}]{wojcieszak2022most}
Wojcieszak, M.; Casas, A.; Yu, X.; Nagler, J.; and Tucker, J.~A. 2022.
\newblock Most users do not follow political elites on Twitter; those who do
  show overwhelming preferences for ideological congruity.
\newblock \emph{Science advances}, 8(39): eabn9418.

\bibitem[{Wojcik and Hughes(2019)}]{wojcik2019sizing}
Wojcik, S.; and Hughes, A. 2019.
\newblock Sizing up Twitter users.
\newblock \emph{PEW research center}, 24.

\bibitem[{Wolfsfeld, Segev, and Sheafer(2013)}]{wolfsfeld2013social}
Wolfsfeld, G.; Segev, E.; and Sheafer, T. 2013.
\newblock Social media and the Arab Spring: Politics comes first.
\newblock \emph{The International Journal of Press/Politics}, 18(2): 115--137.

\bibitem[{Won, Steinert-Threlkeld, and Joo(2017)}]{won2017protest}
Won, D.; Steinert-Threlkeld, Z.~C.; and Joo, J. 2017.
\newblock Protest activity detection and perceived violence estimation from
  social media images.
\newblock In \emph{Proceedings of the 25th ACM international conference on
  Multimedia}, 786--794.

\bibitem[{Yardi and Boyd(2010)}]{yardi2010dynamic}
Yardi, S.; and Boyd, D. 2010.
\newblock Dynamic debates: An analysis of group polarization over time on
  twitter.
\newblock \emph{Bulletin of science, technology \& society}, 30(5): 316--327.

\end{thebibliography}
\end{document}